\documentclass[10pt,conference]{IEEEtran}
\usepackage{dsfont}

\usepackage{color}
\usepackage{setspace}
\usepackage{clrscode}
\usepackage{amsthm}
\usepackage{pict2e}
\usepackage{amsmath}
\usepackage{amssymb}
\usepackage{graphicx}
\usepackage{bbold}
\usepackage{algorithm}
\usepackage{algorithmic}

\usepackage{caption}
\usepackage{subcaption}
\usepackage{mathabx}
\usepackage[ps2pdf,
bookmarks=false,
bookmarksnumbered=false, 
bookmarksopen=false, 
colorlinks=false]{}

\DeclareMathOperator*{\argmax}{argmax}

\newcommand{\Hbf}{\textbf{H}}
\newcommand{\abf}{\textbf{a}}
\newcommand{\vbf}{\textbf{V}}

\newcommand{\sbf}{\textbf{s}}
\newcommand{\rf}{\text{RF}}
\newcommand{\mF}{\mathcal{F}}

\allowdisplaybreaks[2]

\IEEEoverridecommandlockouts
\IEEEpubid{\makebox[\columnwidth]{978-1-7281-4490-0/20/\$31.00 \copyright~2020~IEEE }
\hspace{\columnsep}\makebox[\columnwidth]{ }}

\begin{document}
	\title{Multi-Agent Double Deep Q-Learning  for Beamforming in mmWave MIMO Networks \thanks{This work has been supported in part by the National Science Foundation (CCF-1618615).}}
	
	\author{Xueyuan Wang, M. Cenk Gursoy
		\\Department of Electrical Engineering and Computer Science,
		Syracuse University, Syracuse, NY 13244
		\\Email: xwang173@syr.edu, mcgursoy@syr.edu}

	
	\maketitle
	
	\begin{abstract}
		Beamforming is one of the key techniques in millimeter wave (mmWave) multi-input multi-output (MIMO) communications. Designing appropriate beamforming not only improves the quality and strength of the received signal, but also can help reduce the interference, consequently enhancing the data rate.  In this paper, we propose a distributed multi-agent double deep Q-learning algorithm for beamforming in mmWave MIMO networks, where multiple base stations (BSs) can automatically and dynamically adjust their beams to serve multiple  highly-mobile user equipments (UEs). In the analysis, largest received power association criterion is considered for UEs, and a realistic channel model is taken into account. Simulation results demonstrate that  the proposed learning-based algorithm can achieve comparable performance with respect to exhaustive search while operating at much lower complexity.
	\end{abstract}

	\thispagestyle{empty}

	\thispagestyle{empty}

	\section{Introduction}
	With the rapid growth in data traffic, next-generation wireless
	communication systems are required to provide greater
	throughput to meet higher data-rate demands \cite{2D_DL_HHuang}.
	Under the light of this fact, millimeter wave (mmWave) multi-input multi-output (MIMO) communication has attracted much attention recently.
	In mmWave MIMO networks, beamforming is an effective technique that improves the quality and strength of the received signals by steering the signals generated from an array of transmit antennas to
	an intended angular direction \cite{MIMO_IAhmed}.
	
	 Traditionally, finding the optimal beamforming solution has relied on optimization methods and iterative algorithms, which can lead to high computational complexity and increased delays and thus may not be suitable for real-time implementation \cite{BFNN_WXia}. Additionally,  in a mobile environment, e.g., a network with mobile user equipments (UEs), UEs need to be frequently handed over from one base station (BS) to another, leading to increased control overhead and latency \cite{2D_DL_AAlkhateeb}. With advances in machine learning, there has been increased interest in deep learning and reinforcement learning algorithms to provide low-complexity solutions with low delay. Such learning-based approaches are also regarded as promising for beamforming problems.
	
	\subsection{Related Works}
	As noted above, leveraging recent advances in machine learning, optimal beamforming schemes can be determined in real time with low computational complexity using learning techniques.
	For instance, the authors in \cite{2D_DL_AAlkhateeb} have considered a network where a number of distributed BSs simultaneously serve one mobile UE. The UE ideally transmits one uplink training pilot sequence to all BSs equipped with omni or quasi-omni directional beam patterns, and the deep-learning model leverages the signals to train its neural network. After training, the deep-learning predicts the BS RF beamforming vectors in downlink data transmission.
	\cite{2D_DL_TMakymyuk} has proposed an algorithm that combines three neural networks for performance optimization in massive MIMO beamforming.  In the proposed system, one neural network is trained to generate realistic user mobility patterns, which are then used by a second neural network to produce relevant antenna diagrams. Meanwhile, a third neural network estimates the efficiency of the generated antenna diagrams and returns the corresponding reward to both networks.
	The authors in \cite{2D_DL_WXia} have proposed a deep learning framework for the optimization of downlink beamforming. In particular, the solution is obtained based on convolutional neural networks and exploitation of expert knowledge, such as the uplink-downlink duality and the structure of known optimal solutions.
	In \cite{DNN_BF_XLi}, a neural network architecture is used to jointly sense the millimeter wave channel and design hybrid precoding matrices. The neural network is first trained  in a supervised manner, where a dataset of the mmWave channels and the corresponding RF beamforming/combining matrices are constructed and fed as the input and the target of the neural network, respectively. Then the trained neural network is applied online.

	Reinforcement learning is also shown to be a useful tool for beamforming schemes.
	 For instance, the authors in \cite{2D_RL_FMismar} used deep Q learning algorithm to jointly optimize the beamforming vectors and the transmit power of the BSs, and eventually to maximize the signal-to interference-plus-noise ratio (SINR) of UEs.
	\cite{shafin2019self} presented a deep reinforcement learning framework to optimize MIMO broadcast beams autonomously and dynamically based on users’ mobility patterns or changes in user distribution, which can vary periodically. Using ray-tracing data, deep reinforcement learning engine is first trained offline, and then deployed online for real-time operation. Therefore, whenever the environment changes, the learning engine should be re-trained offline based on the new environment.

In this paper, we consider a general setting and propose a distributed multi-agent double deep Q-learning network (DDQN) solution for beamfoming in mmWave MIMO networks, where multiple BSs serve multiple mobile UEs. In this system, UEs move to different locations at each time, and may be served by different BSs according to the adopted largest received power association criterion. Each BS is a reinforcement learning agent and has its own DDQN. BSs, at each time, can only get  information from the associated UEs, based on which the BSs predict the UEs mobility pattern and choose their beamforming vectors. Hence, the proposed  distributed multi-agent DDQN solution adapts to UEs' mobility. As key novel aspects compared to prior studies, we build a distributed multi-agent learning framework, and the proposed solution requires less feedback from the UEs, and the UEs' feedback is only locally available to the serving BS. Consequently, we have novel designs and algorithms for the learning agents in a more general setting.

\section{System Model}
	In this section, we describe the considered mmWave MIMO network.
	
	\subsection{System Model}
	We consider multi-BS multi-UE MIMO networks, where $J$ BSs are simultaneously serving $K$ mobile UEs. Each BS is equipped with $N_t$ antennas and $N_{\rf}$ RF chains. We assume that the BSs apply analog-only beamforming using networks of phase shifters, and each RF chain is fully-connected with each antenna.
	
	In this analog-only beamforming network, the BS up-converts the data stream to the carrier frequency by passing it through $N_{\rf}$ RF chains. Following this,  the BS uses an $N_t \times N_{\rf}$ RF precoder $\vbf_{\rf}$, which is implemented using analog phase shifters, i.e., with $ |\vbf_{\rf}(a,b)|^2=\frac{1}{N_t}$, to construct the final transmitted signal. Note that $\vbf_{\rf}(a,b) =  \frac{1}{\sqrt{N_t}}e^{j \phi_{ph}}$ where $\phi_{ph}$ is the phase shift angle.
	 For the $k$-th user, the received signal can be modeled as
	\begin{align}
		\textbf{y}_k = \sum_{j=1}^{J} \Hbf_{kj} \vbf_{\rf_j}  \sbf_{kj}  + \textbf{n}_{kj}
	\end{align}
	 where $\sbf_{kj}$ is the data stream vector for the $k$-th UE from the $j$-th BS. We assume that $\text{E}[\sbf_{kj} \sbf_{kj}^H] = P_j \textbf{I}_{N_s}$, and the noise factor $\textbf{n}_{kj} \sim \mathcal{CN} (0,\sigma^2 \textbf{I}_{N_t})$. $\Hbf_{kj}$ is the channel response from the $j^{th}$ BS to the $k^{th}$ UE. $\vbf_{\rf_j}$ is the RF precoder of the $j^{th}$ BS.
	
	\subsection{Channel Model}
	We adopt a channel model with $L$ paths. $L$ is a small number for mmWave communications, and we assume $L=1$ for line-of-sight (LOS) links.  Now, the mmWave MIMO  channel $\Hbf$ can be expressed as
	\begin{align}
		\Hbf=\sqrt{\frac{N_t N_r}{L}}\sum\limits_{l=1}^{L} \frac{\alpha_l}{p_l} \abf_r(\phi^l_r) \abf^H_t(\phi^l_t)
	\end{align}
	where $\alpha_l$ and $p_l$ are the complex gain and the path loss of path $l$, respectively. $\abf_t$ and $\abf_r$ are the array response vectors  at the BS and the UE sides, respectively. $\phi^l_t$ and $\phi^l_r$ are the angles of departure and arrival of the $l^{th}$ path, respectively.
	
	\subsubsection{Array response}
	While the algorithms introduced in  this paper can be applied to arbitrary antenna arrays, we provide the following two illustrative examples of commonly-used antenna arrays.
	For an $N$ element uniform linear array (ULA) on the $y$-axis, the array response vector can be written as \cite{MIMO_OAyach}
	\begin{align}
		\abf_{ULA_y} (\phi) = \frac{1}{\sqrt{N}}\left[1, e^{jkd\sin(\phi)}, ... , e^{j(N-1)kd \sin(\phi)}  \right]
	\end{align}
	where $k = \frac{2\pi}{\lambda}$, $\lambda$ is the wavelength, and $d$ is the inter-element spacing. In the case of a uniform planar array (UPA) in the $yz$-plane with $W$ and $H$ elements on the $y$ and $z$ axes respectively, the array response vector is given by
	\begin{align}
	\abf_{UPA} (\phi, \theta) &= \frac{1}{\sqrt{N}}\Big[ 1, ... , e^{jkd(m\sin(\phi) \cos(\theta)+ n \sin(\phi)\sin(\theta))}, \notag \\
	 &... , e^{jkd((W-1)\sin(\phi) \cos(\theta)+ (H-1) \sin(\phi) \sin(\theta))}  \Big]
	\end{align}
	where $0<m<W$ and $0<n<H$ are the $y$ and $z$ indices of an antenna element respectively and the antenna array size is $N=WH$. In this paper, we primarily concentrate on ULA.
	
	\subsubsection{Path Loss}
	Link between a UE and a BS can be either LOS or non-LOS (NLOS). The path loss model is formulated as
	\begin{align}
	p_l(r)=
	\begin{cases}
	\kappa^{los} r^{\alpha^{los}}(r) & \text{with prob. } \quad p^{los}(r)\\
	\kappa^{nlos} r^{\alpha^{nlos}}(r) &  \text{with prob. }  \quad p^{nlos}(r)=(1-p^{los}(r))
	\end{cases}
	\end{align}
	where $r$ is the two-dimensional distance between the UE and BS. $\alpha^{los}, \alpha^{nlos}$ are the path loss exponents for LOS and NLOS links, respectively, $\kappa^{los}, \kappa^{nlos}$ are the intercepts of the LOS and NLOS path loss formulas, respectively, and $p^{los}(r)$ is the probability that the link has a LOS transmission at distance $r$.
	
	Following the 3GPP standards described in \cite{3GPP_36828}, we express the probability of LOS link between the BSs and the UEs as
	\begin{align}
	p^{los}(r)= \min(18/r,1)\times(1-\exp(-r/63))+\exp(-r/63).
	\end{align}
	
	Shadowing is also taken into account in the channel model, and is modeled as a log-normal random variable, i.e., $10 \log v \sim  \mathcal{N}(\mu_v, \sigma_v^2)$ with $\mu_v$ and $\sigma_v^2$ being the mean and variance of the channel power under shadowing, respectively.

	\subsection{Achievable Rate}
	In the multi-BS multi-UE mmWave MIMO network, beside the serving BS, other BSs inflict interference to the UEs.  We can express the rate of the $k$-th UE when associated with the $j$-th BS  as \cite{2D_RL_FMismar} \cite{jiang2016joint} \cite{MIMO_AAlkhateeb}
	\begin{align}
		R_k=\log_2 \left(1+ \frac{P_j |\Hbf_{kj} \vbf_{\rf_j} \vbf^H_{\rf_j} \Hbf^H_{kj}| }{\sigma^2 + \sum\limits_{i=1,i\neq j}^{J}P_i|\Hbf_{ki} \vbf_{\rf_i} \vbf^H_{\rf_i} \Hbf^H_{ki} | }  \right)
	\end{align}
	where superscript $H$ and $|\cdot|$ denote the conjugate transpose and the determinant of a matrix, respectively.
	 Then, the sum-rate of all UEs is $R_{sum} =\sum\limits_{k=1}^{K} R_k$.
	
	
   Our goal is to maximize the sum-rate of all UEs, and therefore we consider the following optimization problem:
	\begin{align}
	&\max\limits_{\vbf_{\rf_j}, j\in J}  \hspace{0.8in} \sum\limits_{k=1}^{K} R _k  \\
	&\text{subject to } \qquad |\vbf_{\rf_j}(a,b)|^2 = \frac{1}{N_t} \forall j, \tag{8a}
	\end{align}
	where $\vbf_{\rf_j}$ is the analog beamforming vector of the $j$-th BS.

\section{Distributed DDQN for mmWave MIMO networks}
	In this section, we first describe the generalized form of DDQN, and then introduce the proposed distributed multi-agent DDQN in detail.
	\subsection{Generalized form of DDQN} 	

	In reinforcement learning, an agent dynamically interacts with an unknown environment  $\mathcal{E}$, and makes sequential decisions.  At each time step, the agent is in  a state, $s_t \in \mathcal{S}$, selects an action, $a_t \in \mathcal{A}$, then receives an immediate scalar reward $r_t\in \mathcal{R}(s_t, a_t)$, and transitions to next state $s_{t+1}$. The cumulative discounted reward, $\mathbb{R}_t$, at time step $t$, is defined as $\mathbb{R}_t = \sum\limits_{k=0}^{\infty} \gamma^k r_{t+k}$, where $\gamma \in (0,1]$ is the reward discount factor, which balances the importance of immediate and future rewards. Q-learning is one of the most widely used algorithms for reinforcement learning. In Q-learning, the agent learns a state-action value function $Q^{\pi}(s,a)$, which is defined as the expected cumulative reward when the agent takes action $a_t$ in state $s_t$ following its policy $\pi$, and can be expressed as $Q^{\pi}(s,a) = E[\mathbb{R}_t|s,a]$. The optimal state-action value function is denoted as $Q^*(s,a) = \max_{\pi} Q^{\pi}(s,a)$, which satisfies the Bellman optimality equation \cite{RL_MIT} $Q^*(s,a) = E[r_t + \gamma \max_{a'} Q^*(s_{t+1},a')|s,a]$. The goal is to find an optimal policy $\pi^*$ that maximizes the optimal action value function $\pi^*(s) = \argmax_a Q^*(s,a)$. The Q-learning update rule in its general form is given by
	\begin{align}
		Q(s,a) \leftarrow  Q(s,a)+ \alpha[r_t + \gamma \max_{a'} Q(s_{t+1},a') - Q(s,a)]
	\end{align}
	where $\alpha$ is a scalar step size.
	
	 In problems with large state and action spaces,  we can learn a parameterized value function $Q(s,a;\xi)$ \cite{van2016deep}.  A deep Q network consists of  an online deep Q-learning phase and an offline deep neural network (DNN) construction phase, which is used to learn the value function $Q(s,a;\xi)$ where $\xi$ denotes the set of the parameters of the DNN. Generally, the parameter set $\xi$ can be optimized by minimizing the following loss function \cite{DDQN_PLv}
	 \begin{align}
	 	L_t(\xi_t) = E[y_t - Q(s_t,a_t;\xi_t)]^2
	 \end{align}
	 where $y_t = r_t +\gamma \max_{a} Q(s_{t+1},a; \xi^-_t)$  is the objective function, and $\xi^-_t$ is copied every $\tau$ steps from the $\xi_t$.
	
	 The max operator in standard Q-learning and DQN uses the same values both to select and to evaluate an action, which increases the probability to select overestimated values and results in overoptimistic value estimates. In DDQN, the objective function can be written as  \cite{van2016deep}
	 \begin{align}
	 	y_t = r_t +\gamma Q(s_{t+1}, \argmax_a Q(s_{t+1}, a;\xi_t); \xi^-_t),
	 \end{align}
	 where $\xi_t$ is used for action selection, and $\xi^-_t$ is used to evaluate the value of the policy, and $\xi^-_t$ can be updated symmetrically by switching the roles of $\xi_t$ and $\xi^-_t$.
	
	\subsection{Distributed multi-agent DDQN }
	\begin{figure}
		\centering
		\includegraphics[width=0.5 \textwidth]{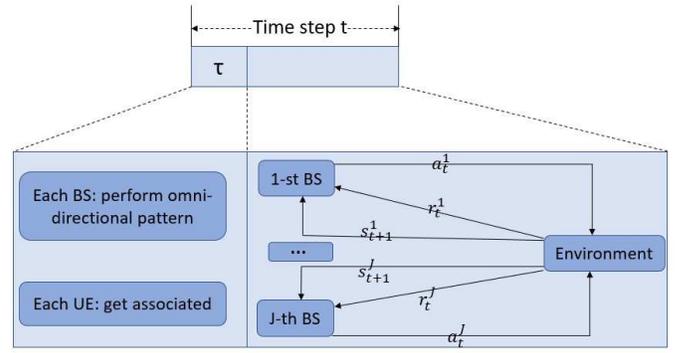} \\
		\caption{\small An illustration of the procedure in one time step. \normalsize}
		\label{figure_ddqn}
	\end{figure}
	The algorithm of the distributed multi-agent DDQN for mobile UEs is described in this subsection. Before providing the definitions of states, actions, and rewards,  the environment should be introduced first. In the environment $\mathcal{E}$, BSs are located with a certain distance in between.  UEs move from different initial locations in different directions with different speeds.  In each episode, UEs do not change their speed and direction until they move out of the coverage region of all BSs,  and the episode ends. In each episode, at each time step, UEs move to new locations. We assume that at the beginning of each time step, in a very short time slot $\tau$, each BS uses an omni-directional antenna pattern, and transmits pilot signals to all UEs. Therefore,  each UE receives pilot signals from all BSs, and chooses to associate with the BS providing the strongest pilot signal, and computes the immediate rate under the omni-directional pattern, and sends it to the serving BS, indicating the association condition with itself. It is assumed that each UE is equipped with a memory containing prior information (e.g.,  the location history, the omni-directional rate in the last $T_m$ time steps), and each UE can send this information to the associated BS as well. Subsequently,  the BS performs beamforming according to our algorithm, and serves the associated UEs. Then the UEs learn the immediate rate achieved with beamforming, and feed this information back to the serving BS. The procedure in one time step is illustrated in Fig. \ref{figure_ddqn}.

    	\begin{algorithm}
    	\caption{Distributed multi-agent DDQN}
    	\label{ddqn}
    	\begin{algorithmic}[1]
    		\REQUIRE
    		\STATE Initialize replay memory $D^j$ of DDQN,  $j\in J$, to capacity N for every BS.
    		\STATE Initialize online network with random parameter $\xi^j$, $j\in J$.
    		\STATE Initialize target nework with parameter $\xi^{j-} = \xi^j$, $j\in J$.
    		\ENSURE
    		\FOR{episode = 0: total episode}
    		\STATE Reset environment $\mathcal{E}$.
    		\STATE Initialize $s^j_1$ for $j\in J$.
    		\FOR{$t=1:T$}
    		\FOR{$j=1:J$}
    		\STATE Obtain association condition and the sum-rate of all associated UEs $R^{oj}_t$ by association procedure.
    		\STATE Sample $c$ from Uniform (0,1)
    		\IF{ $c\leq \epsilon$ }
    		\STATE Select an action (beamforming vector index) randomly from the codebook $\mF$.
    		\ELSE
    		\STATE Select the action $a^j_t = \argmax_{a} Q^*_j(s^j_t,a;\xi^j_t)$.
    		\ENDIF
    		\STATE Execute action $a^j_t$, i.e. apply the selected beamforming vector on the antenna arrays of $j$-th BS.
    		\STATE Observing the resulting state $s^j_{t+1}$ and the immediate sum-rate of all associated UEs $R^j_t$.
    		\STATE Compute the immediate reward $r^j_t$, i.e. $\frac{R^{oj}_t}{R^j_t} \times 100 \% $.
    		\STATE Store the experience tuple  ($s^j_t, a^j_t, r^j_t,s^j_{t+1}$) in $D^j$.
    		\STATE Sample random minibatch of experience ($s^j_{\tau}, a^j_{\tau}, r^j_{\tau},s^j_{\tau+1}$) from $D^j$.
    		\STATE Update \\
    		$\small y^j_{\tau}=$\\$ \begin{cases}
    		r^j_{\tau} \qquad\text{if episode terminates at step } \tau+1\\
    		r^j_{\tau}+\gamma Q(s^j_{\tau+1}, \argmax_{a'} Q(s^j_{\tau+1}, a';\xi^j_{\tau}); \xi^{j-}_{\tau}) \\ \hspace{1in} \text{otherwise}
    		\end{cases}$
    		\STATE Perform a gradient descent step on $(y^j_{\tau}-Q(s^j_{\tau}, a;\xi^j_{\tau}))^2$ with respect to the network parameters $\xi^j_{\tau}$
    		\STATE For every $N_n$ steps reset target network parameter $\xi^{j-} = \xi^j$
    		\ENDFOR
    		\ENDFOR
    		\ENDFOR
    	\end{algorithmic}
    \end{algorithm}

	In the  multi-agent DDQN model, each BS is an agent and the state, action, and reward tuple of the $j$-th BS is denoted by $(s^j_t,a^j_t,r^j_t)$ . These states, actions and rewards are described in detail below:
	\subsubsection{State}
	
	At each time step, each BS serves multiple UEs, and is able to obtain the information in the memory of each UE. This information of all associated UEs constitute as state at time $t$, noted as $s^j_t$ for the $j$-th BS. Assume that the length of history in each memory is $T_m$, and there are $K$ UEs in this network, then the state $s^j_t= [m^0_{t-T_m},m^0_{t-T_m+1},...,m^0_{t-1},...,m^K_{t-1} ]$. It is worth noting that for UEs not associated with the $j$-th BS, $m^*_{t-*}$ is set to be 0. In other words, the BSs only need information from the associated UEs.  In this paper, we denote the rate achieved with the omni-directional antenna pattern as $m^*_{t-*}$ . If the location information of each UE is also available, we denote the omni-directional pattern rate and the location information together as state $s^j_t$ . In Section IV, we provide performance results with and without location information.
	\subsubsection{Action}
	At each time step, each BS chooses an analog beamforming vector $\vbf_{\rf}$. Due to the constraints on the RF hardware, such as the availability of only certain quantized angles for the RF phase shifters, the analog beamforming vectors can take only certain values. Hence, finite-size codebooks for the candidate beamforming vectors are needed. In practice, the beamforming vectors are spatial matched filters for the single-path channels \cite{MIMO_AAlkhateeb}. Thus, they have the same form of the array response vector and can be parameterized by a simple angle.  While the algorithm in this paper can be applied to arbitrary finite-size codebooks, we adopt  the codebook, denoted by $\mF$, consisting of the steering vectors $\abf_t(\phi_Q)$ where $\phi_Q$ is the quantized angle. The beamforming vector index in the codebook is defined as the action $a^j_t$.
	
	\subsubsection{Reward}
	 As noted before, at the beginning of each time step, each BS learns the immediate rate, $R^{oj}_t$, provided to all associated UEs with the omni-directional antenna radiation pattern. After performing beamforming, each BS also learns the immediate rate $R^j_t$ achieved with beamforming.  We regard the ratio of two rates, $\frac{R^j_t}{R^{oj}_t}$ as the immediate reward $r^j_t$ of the $j$-th BS when this BS takes action $a^j_t$ in state $s^j_t$.
	
	Setting the analog beamforming vector codebook $\mF$, the optimization problem can be reformulated as
	\begin{align}
	&\max\limits_{\vbf_{\rf_j}[t], j\in J}  \hspace{0.3in} \sum\limits_{k=1}^{K} R _k[t]  \\
	&\text{subject to } \qquad \vbf_{\rf_j}[t] \in\mF ,  \forall j, \tag{12a}
	\end{align}
	where $\vbf_{\rf_j}[t]$ is the analog beamforming vector of the $j^{th}$ BS at time $t$.
	

\section{Performance Evaluation}
	In this section, we provide simulation results and evaluate the performance of the proposed multi-agent DDQN for mmWave MIMO beamforming.
	
	\subsection{Environment Setting}
\begin{figure}
		\centering
		\includegraphics[width=0.45 \textwidth]{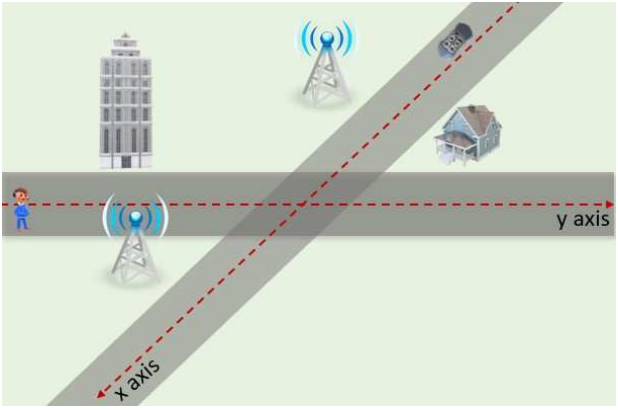} \\
		\caption{\small An illustration of the simulation environment. \normalsize}
		\label{figure_street}
	\end{figure}

	The considered environment is illustrated in Fig. \ref{figure_street}. As shown in the figure, we have two intersecting streets. The BSs are located along the streets, while the UEs are moving from the beginning of either street in either direction. UEs move at random speeds. The length of each street is set as 100m, starting from -50m to 50m on each axis. The width of the road is set as 8m, i.e., (-4m, 4m). The speed of the UEs can be between 2 to 5 m/s. The locations of BSs are set at the coordinates of [(5,-5),(-25,-5)]. In addition, the number of antennas at the base stations and the UEs are set, respectively, as $N_t = 16, N_r=1$. 	Finally, the number of RF chains is $N_{\rf} = 2$.

	\subsection{Hyperparameters}
	In our experiment, we construct the DDQN via three-layered neural networks using Adam optimizer to evaluate the gradient descent of the evaluated and target networks. Our input size depends on the number of UEs in the environment, i.e., the input size is $KT_m$ when $K$ is the number of UEs and $T_m$ is the length of historic/prior information at of each UE, and is set to be 8 in the simulation. The output size should be the size of the action notebook $\mF$. The number of neurons of the three layers are $12K$, $8K$ and 8, respectively. The discount factor is 0.95, batch size is 32, and the learning rate for two BSs are 0.0001 and 0.005, respectively. We also use $\epsilon$-greedy policy and the maximum value of $\epsilon$ is 0.9 and the minimum is 0.1.

	\begin{figure}
	\centering
	\begin{minipage}{0.5\textwidth}
		\centering
		\includegraphics[width=1 \textwidth]{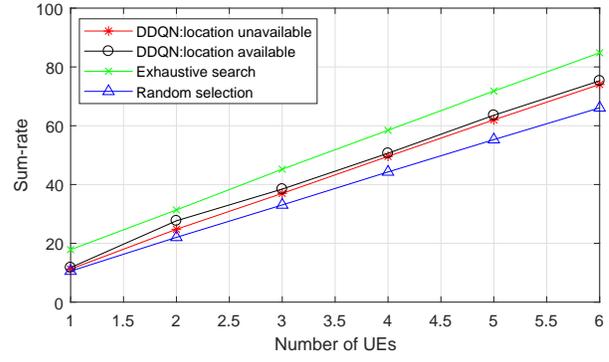} \\
		\subcaption{\scriptsize Single Base Station. }
	\end{minipage}
	\begin{minipage}{0.5\textwidth}
		\centering
		\includegraphics[width=1\textwidth]{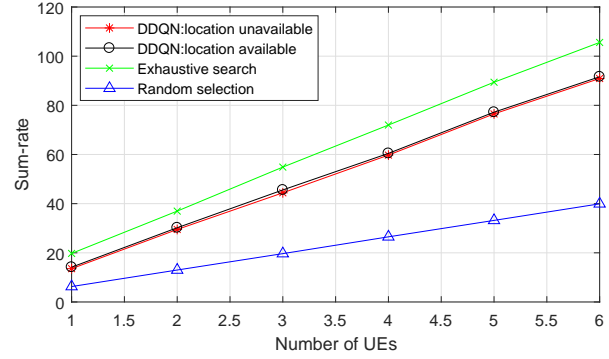}
		\subcaption{\scriptsize Two Base Stations.}
	\end{minipage}
	\caption{\small Sum-rate as a function of number of UEs for different number of BSs.}
	\label{figure_1BS2BS}
\end{figure}
	\begin{figure}
		\centering
		\includegraphics[width=0.5 \textwidth]{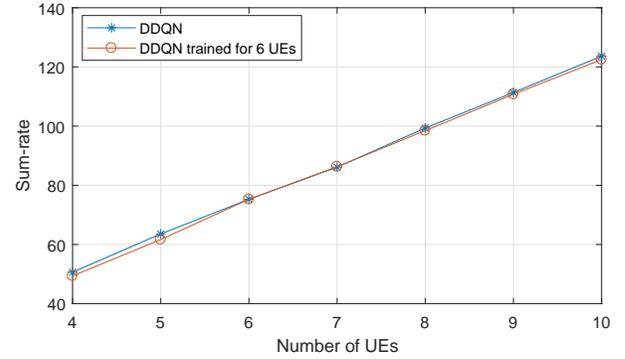} \\
		\caption{\small A performance comparison of testing results, when the DDQN is trained for the exact number of UEs and the DDQN is trained for 6 UEs. The number of BS is one. \normalsize}
		\label{figure_com}
	\end{figure}
	
	\subsection{Experiment results}
	Fig. \ref{figure_1BS2BS} plots the sum-rate as a function of the number of UEs for different number of BSs. Results from exhaustive search among all possible beamforming directions and also  random selection are provided as two benchmark results. Note that the exhaustive search requires perfect channel state information (CSI), and in random selection the BSs randomly choose actions from the codebook. Figs. \ref{figure_1BS2BS}(a) and \ref{figure_1BS2BS}(b) (in which we consider the cases of a single BS and two BSs, respectively) show that DDQN can achieve better performance than random selection, and comparable results with exhaustive search which has high computational complexity and incurs potentially large delays especially in mobile scenarios.  In addition, in both figures, we provide the performance curves when the location information is available and unavailable. The performance with location information given is only slightly better, indicating that such information is not critical for our algorithm. Furthermore, when the number of UEs increases, the sum-rate grows almost linearly, demonstrating that the proposed algorithm can handle multiple UEs without much decrease in the data rate experienced at each UE. On the other hand, when we compare Figs. \ref{figure_1BS2BS}(a) and \ref{figure_1BS2BS}(b), we notice that for the same number of UEs, two BSs can provide higher sum-rate than in the case of a single BS. Even though there is interference when there are multiple BSs, two BSs can provide larger coverage, leading to higher SINR levels and rates with effective beamforming. We also notice that random selection performs rather poorly especially in  the presence of two BSs.
	
	Fig. \ref{figure_com} displays the sum-rate performances during testing, when the DDQN is trained for the exact number of UEs used in the tests and also when  DDQN is trained considering 6 UEs regardless of how many UEs we have in the test period. When the DDQN is trained for 6 UEs, if the real number of UEs is less than 6 in the testing, we need to do zero padding in the input of the DDQN; if the actual number of UEs is more than 6, we just randomly choose 6 UEs and make a decision. From Fig. \ref{figure_com}, we observe that when the DDQN is trained for the actual number of UEs , the testing performance is slightly better than that of the other case, indicating that the pre-trained model does not need to be restricted to a certain number of UEs and can be applied to scenarios in which the number of UEs in the test is different from that in the training period.

\section{Conclusion}
	In this paper, we have  proposed a multi-agent DDQN algorithm for beamforming in mmWave MIMO networks. Largest received power association criterion has been considered for BS association of the UEs. BSs act as reinforcement learning agents, and according to the limited information obtained from the associated UEs, they automatically and dynamically adjust their beams to improve the received power of the associated UEs. Via simulations, we have demonstrated that the proposed algorithm can achieve comparable network performance with respect to exhaustive search, and better performance than the random selection, which leads to especially poor performance when there are multiple BSs. We have noted that location information is not critical in our algorithm. In addition, the pre-trained model is not restricted to the same number of UEs as in the testing phase and can be applied to multiple testing scenarios with different number of UEs. 3D beamforming in mmWave MIMO networks remains as an interesting future research direction in which DDQN can also be employed.

	\bibliographystyle{IEEEtran}
	\bibliography{cluster_milli}
	
\end{document}